\documentclass[mmnp]{edpsmath}
 \usepackage{xcolor}
 \usepackage{soul}

 \usepackage{graphicx}
 \usepackage{dcolumn}
 \usepackage{bm}
 \usepackage{changepage}
 \usepackage{hyperref}
 \usepackage{cite}
 \usepackage{comment}
 \usepackage{ulem}
\begin{document}
\title{A model for seagrass species competition: dynamics of the symmetric case}\thanks{Corresponding author \email{pablo@ifisc.uib-csic.es}}
\author{Pablo Moreno-Spiegelberg}\address{IFISC (CSIC-UIB), Instituto de F\'{\i}sica Interdisciplinar y Sistemas Complejos,  E-07122 Palma de Mallorca,  Spain}
\author{Damià Gomila}\sameaddress{1}
\date{\today}
\begin{abstract} 
We propose a general population dynamics model for two seagrass species growing and interacting in two spatial dimensions. The model includes spatial terms accounting for the clonal growth characteristics of seagrasses, and coupling between species through the net mortality rate. We consider both intraspecies and interspecies facilitative and competitive interactions, allowing  density-dependent interaction mechanisms. Here we study the case of very similar species with reciprocal interactions, which allows reducing the number of the model parameters to just four, and whose bifurcation structure can be considered the backbone of the complete system. We find that the parameter space can be divided into ten regions with qualitatively different bifurcation diagrams. These regimes can be further grouped into just five regimes with different ecological interpretations. Our analysis allows the classifying of all possible density distributions and dynamical behaviors of meadows with two coexisting species.
\end{abstract}
\subjclass{92D25,35B36,35B32,35K55}
\keywords{Population dynamics, competition, facilitation, Allee effect, seagrasses}
\maketitle

\section{Introduction}

Seagrass meadows are key to marine coastal ecosystems \cite{Costanza1997}. They provide food, protection, and structural support to many marine species \cite{Beck2001}. Moreover, seagrass meadows are an important sink of carbon dioxide \cite{Duarte2005}, protect the coastline against strong waves \cite{Fonseca1992,Sanchez-Gonzalez2011}, and contribute to nutrient sedimentation. From a socioeconomic point of view, seagrass ecosystems support fishing and human development.
During the last decades, a decline in seagrass beds associated with trawling, pollution,  global warming, or competition with invasive species, among other anthropogenic effects, has been observed \cite{Orth2006,Hughes2009,Waycott2009}. 
Preventive, palliative, and restoration measures must be taken to reduce the consequences of this declining \cite{Halpern2007,Ryabinin2019}. Not only seagrasses are in danger, but about half of the marine ecosystems have also been identified as strongly affected by multiple anthropogenic drivers \cite{Halpern2008}. No wonder UN has declared with urgency 2021-2030 as the “Decade of Ocean Science for Sustainable Development” as well as the “Decade of Ecosystem Restoration”. 

Dynamical models provide a framework to study the meadow receding process and to understand the mechanisms that govern the ecosystem dynamics. This can be used to estimate the resilience and alert about the proximity of tipping points, after which vegetation systems collapse. Furthermore, they can also be used to make predictions about the evolution of the meadows under different scenarios. This provides useful information to take decisions in ecosystems management. 

Two different approaches have been used to study the dynamics of seagrass meadows. The first is based on microscopic agent based models where information on each plant shoot (and apex) is explicitly computed. The dynamics are defined in these systems as a markovian process at shoot level, where apices grow and/or branch, generating new shoots and apices, and both die with a given rate \cite{Sintes2005,Sintes2006a}. 
The second approach is based on macroscopic models where only spatial densities are considered \cite{Ruiz-Reynes2017,Ruiz-Reynes2020c}. In these models, the evolution of plant density is described by a system of partial differential equations (PDEs). Even if the macroscopic models lack information on individual shoots and the rhizome network, they are computationally more efficient to study large systems. Furthermore, bifurcation analysis can be applied to PDEs, providing analytical information about instabilities and tipping points under changing conditions.

Interaction between species is a relevant mechanism in seagrasses dynamics.
While some species of seagrasses coexist in space creating mixed meadows, others arrange in separated monospecies beds with interfaces. Some species have been seen in both arrangements for different conditions,  suggesting some kind of transition between these behaviors. Interspecies interaction is then key in determining the evolution of ecosystems with invasive species. In a global change scenario like the one we are currently experiencing, the interaction between native species with different responses to the new conditions, e.g. due to global warming, can also determine the evolution of the ecosystems \cite{Collier2011,Savva2018}.
Introducing interspecific interactions to current seagrasses models is necessary to study this process. So far, in the context of seagrass dynamics, interactions between species have only been studied in microscopic models \cite{Llabr2022a,Llabr2022b}. The addition of interspecies interaction in macroscopic models of seagrasses is, so far, unexplored.

 In this work, we present a generalization of a single species seagrass macroscopic model \cite{Ruiz-Reynes2020c} considering local interspecies interaction. Furthermore, we study in detail the bifurcation diagram of the symmetric case, where the two species are similar and the interaction between them is reciprocal. This simple scenario captures the backbone of the  general model and, despite its simplicity, it gives a remarkable variety of scenarios with complex behaviors. These scenarios can be related to biotic interactions between species, while the transitions between them are mediated by abiotic (environmental) changes in the mortality rate. 

\section{The Model}

In \cite{Ruiz-Reynes2020c}, a simple model to describe meadows of clonal-growth plants was proposed. In that work the evolution of the plant density $n(\vec{r},t)$ is described by the following partial differential equation:
\begin{align}
    \label{Single_PDE}
    \partial_{t} n= -n \omega (n) + d_{0} \nabla^2 n +d_{1}((\nabla^2 n)n+||\nabla n||^2)
\end{align}
where $\omega(n) =-\omega_b(n)+\omega_d(n)$ is the net death rate, being $\omega_b>0$ the branching rate and $\omega_d>0$ the death rate, in principle, both density dependent. The elongation of the rhizome of clonal plants combined with the branching lead to an effective diffusion with coefficient $d_0$ and to a nonlinear diffusion with coefficient $d_1$. Additionally, a gradient squared term with coefficient $d_1$, characteristic of clonal growth, appears also in the model \cite{Ruiz-Reynes2020c}.  

To describe a two species system, using Eq.~(\ref{Single_PDE}), we couple two different vegetation density fields through the mortality term to describe both intraspecific and interspecific interactions:
\begin{align}
    \label{Double_PDE}
    \partial_{t} n_{i}= n_{i} Q_{i} [\vec{n}] + d_{i0} \nabla^2 n_{i} +d_{i1}((\nabla^2 n_{i})n_{i}+||\nabla n_{i}||^2)
\end{align}
where we consider local interactions only in the net mortality term, given by a quadratic polynomial:
\begin{align}
Q_{i}(\vec{n})=&-\omega_i  +\vec{a_{i}}\cdot\vec{n}-b_{i1}^2n_{1}^2 -b_{i2}^2n_{2}^2-b_{i3}n_{1}n_{2}
\label{mortality1}
\end{align}
where $\omega_{i}$ is the net mortality of species $i$ in absence of other plants, $\vec{a_{i}}=(a_{i1},a_{i2})$, and $\vec{n}=(n_1,n_2)$. $a_{ii}$ and $a_{ij}$ $(i \neq j)$ are the slopes of the linear change in the net mortality rate due to intraspecific and interspecific interactions respectively. A term $a_{ij}>0$ describes a facilitative interaction for moderate densities while $a_{ij}<0$ describes a competitive interaction. The quadratic terms $b_{ij}>0$ are saturation parameters that always describe competitive interactions for high plant densities, acting as a carrying capacity and giving an upper bound to plant density. We consider the cross saturation term as $b_{i3}=2b_{ii}b_{ij}$ for $i\neq j$, simplifying the mortality term to a parabolic form:
\begin{align}
Q_{i}(\vec{n})=&-\omega_i  +\vec{a_{i}}\cdot\vec{n}-(\vec{b_{i}}\cdot\vec{n})^2.
\label{mortality2}
\end{align}
This way, considering equal interspecific and intraspecific interactions the mortality term is a function of the total density only, i.e. the sum of the densities of both species $n_1+n_2$, as expected if $n_1$ and $n_2$ were the same species.

The local part of Eq. (\ref{Double_PDE}) corresponds to a generalized Lotka-Volterra equation \cite{Brenig1988,Hernandez-Bermejo1997} with up to quadratic terms in the mortality rate (\ref{mortality2}), in both inter and intraspecies interactions. The use of these nonlinear interactions is supported by both theoretical and field observations. Specifically, recent studies have shown that interspecific plant-plant facilitation is density dependent and it has a single maximum for intermediate densities \cite{LeRoux2010}. Also, monospecific seagrass meadows show an abrupt collapse of the plant population for small increases of a stressor above a given critical value \cite{Mayol2022a,McGlathery2013}, which indicates the presence of tipping points in the system. Both behaviors need at least up to quadratic nonlinear terms in (\ref{mortality2}) to be properly described.

The obtained model is versatile and can represent species with different growth dynamics. It also allows a flexible representation of the different interactions between plants, such as competition, mutualism, amensalism, or parasitism. Additionally, the model is easily scalable to more than two species, making it a useful tool for studying multispecies seagrass meadows dominant in tropical climates. The plasticity of the model allows then for a comprehensive understanding of the complex interactions within ecosystems.

\section{The symmetric case}

In this section, we consider in detail the simplified case in which both plants are similar and have symmetric interactions, in such a way that the mortality and the intraspecies and interspecies terms are the same for both species, greatly reducing the number of parameters. This implies reciprocal interactions, i.e. mutualism and competition are the only possible relationships. For this situation, $\omega_1=\omega_2:=\omega$, $a_{11}=a_{22}:= a_1$, $a_{12}=a_{21}:= a_2$, $b_{11}=b_{22}:= b_1$, $b_{12}=b_{21}:= b_2$, $d_{10}=d_{20}:=d_0$, and $d_{11}=d_{21}:=d_1$. Notice that, in this symmetric case, $Q_1(n_1,n_2)=Q_2(n_2,n_1) \equiv Q(n_1,n_2)$.

Considering low-density intraspecies facilitation (i.e. $a_1>0$), the equations can be reduced to an adimensional form through the change of variables
\begin{align}
    n'_1=\frac{b_1^2}{a_1}n_1 && n'_2=\frac{b_1^2}{a_1}n_2 && t'=\frac{a_1^2}{b_1^2} t && \vec{r'}=\frac{a_1}{b_1\sqrt{d_0}}\vec{r}, 
\end{align}
and using the following rescaled parameters
\begin{align}
    \omega'=\frac{b_1^2}{a_1^2}\omega && \alpha=\frac{a_2}{a_1} && \beta =\frac{b_2}{b_1} && \delta = \frac{a_1}{b_1^2}\frac{d_1}{d_0}.
\end{align}
Dropping the primes, Eqs. (\ref{Double_PDE}) become:
\begin{align}
    \dot{n_1}=&n_1Q(n_1,n_2) + \nabla^2 n_1 +\delta ((\nabla^2 n_1)n_1+||\nabla n_1||^2) \nonumber\\
    \dot{n_2}=&n_2Q(n_2,n_1) + \nabla^2 n_2 +\delta ((\nabla^2 n_2)n_2+||\nabla n_2||^2)
    \label{Eq:Sym_PDE}
\end{align}
where
\begin{align}
Q(n_1,n_2)=-\omega  +n_1+\alpha n_2-(n_1 +\beta n_2)^2.
\end{align}
The new parameter $\omega$ is proportional to the net mortality of plants in the absence of interactions. 
We consider it depends on abiotic factors, i.e. it changes with the environmental conditions. Parameters $\alpha$ and $\beta$ give the ratio between interspecific and intraspecific interactions. Finally, $\delta$ is a parameter proportional to the ratio between nonlinear and linear diffusion. In this work, we assume the parameters $\alpha$, $\beta$, and $\delta$ not to depend on abiotic factors, and to be determined by the characteristics of the interacting species. 

\begin{figure}
\centering
\includegraphics[width=0.5\textwidth]{ 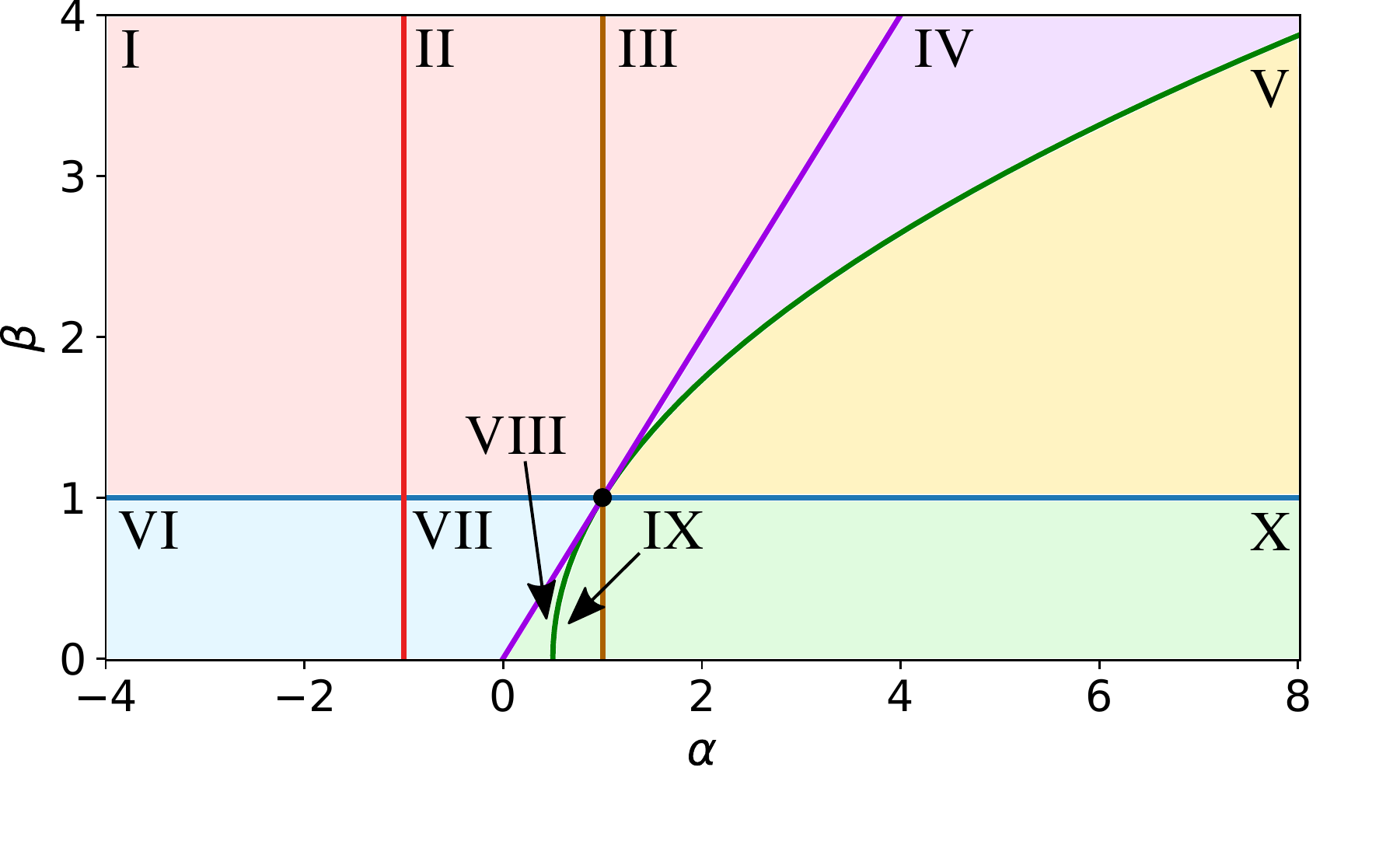}
\caption{Projection of the full phase diagram on the ($\alpha,\beta$) plane. Lines represent codimension-2 bifurcations and singular points. These lines divide the interaction parameter plane ($\alpha, \beta$) in ten different regions with qualitatively unique bifurcation diagrams as a function of $\omega$, labeled with roman numerals. These regions are grouped into 5 ecological cases: competition exclusion shaded in pink (regions I, II, and III); dynamic coexistence shaded in purple (region IV); low-density coexistence, shaded in yellow (region V); high-density coexistence, shaded in blue (regions VI and VII), and mutualism, shaded in green (regions VIII, IX and X).
The red line represents the projection of the codimension-2 bifurcation where $SN_S$ and $T_0$ cross ($\alpha=-1$); the brown line where $Pitch$ and $T_0$ cross; the purple line where $Pitch$, $SN_{S}$ and $Hopf$ converge; and the green curve where $T$, $SN$ and $Hopf$ converge. Finally, the blue line represents a singular case, $\beta=1$, where the value of $\omega$ at which $Pitch$ and $T$ take place diverges to $-\infty$.}
\label{Fig:cod_2_phase_diag}
\end{figure}

Throughout this work, we fix $\delta=0.5$ and use $\alpha$ and $\beta$ as the parameters characterizing the species, and $\omega$ as the control parameter whose variations reflect changes in the environment. For fixed biotic parameters ($\alpha$, $\beta$), a change in the value of $\omega$ can qualitatively modify the behavior of the system by crossing different bifurcation points. Advancing results to be discussed in detail later, we find that the parameter space ($\alpha$, $\beta$) can be partitioned into ten different regions (see Fig.~\ref{Fig:cod_2_phase_diag}), in each of which the bifurcation diagram as a function of $\omega$ is qualitatively different from the others. These ten regions in the parameter space can be further grouped into five different cases (color shaded regions in Fig.~\ref{Fig:cod_2_phase_diag}), each with a different ecological interpretation.   

\subsection{Homogeneous steady solutions and their bifurcations}
\label{Section: Homogeneous steady solutions}

\begin{table*}[t]
 \caption{Homogeneous steady states of Eq. (\ref{Eq:Sym_PDE}).}
    \label{tab:HSS}
    \begin{tabular}{|c |c |c |}
        \hline
        Label &Name & Value ($n_1$,$n_2$) \\ \hline

        $P_0$ & Bared state/unpopulated &  $(0,0)$ \\ \hline
        $P^h$ & High populated monospecific  &  $(0,\frac{1+ \sqrt{1-4\omega}}{2});(\frac{1+ \sqrt{1-4\omega}}{2},0)$ \\ \hline
        $P^l$ & Low populated monospecific  & $(0,\frac{1- \sqrt{1-4\omega}}{2});(\frac{1- \sqrt{1-4\omega}}{2},0)$ \\ \hline
        $P_{S}^{h}$ & High populated symmetric mixed  & $\frac{1+\alpha+\sqrt{(1+\alpha)^2-4\omega(1+\beta)^2}}{2(1+\beta)^2}(1,1)$ \\ \hline
        $P_{S}^{l}$ & Low populated symmetric mixed  & $\frac{1+\alpha-\sqrt{(1+\alpha)^2-4\omega(1+\beta)^2}}{2(1+\beta)^2}(1,1)$ \\ \hline
        $P_{A}$ & Assymetric mixed & $\left( \frac{1-\alpha}{2(1-\beta^2)} \pm \sqrt{\frac{\omega_P -\omega}{(1-\beta)^2}},\frac{1-\alpha}{2(1-\beta^2)} \mp \sqrt{\frac{\omega_P -\omega}{(1-\beta)^2}} \right)$ \\ \hline
    \end{tabular}
\end{table*}

\begin{table*}[t] 
\caption{Local bifurcations of the HSS.}
    \label{tab:bif}
    \begin{tabular}{|c |c |c |} 
        \hline
        Label &Name & Critical point $\omega_c$ \\ \hline

        $T_0$ & Degenerate bared state transcritical &  $0$ \\ \hline
        $SN$ & Monospecific Saddle Node  &  $0.25$ \\ \hline
        $SN_S$ & Symmetric mixed Saddle Node  & $\frac{(1+\alpha)^2}{4(1+\beta)^2}$ \\ \hline
        $T$ & Monospecific transcritical & $\frac{(1-\alpha)(\alpha-\beta^2)}{(1-\beta^2)^2}$ \\ \hline
        $Pitch$ & Pitchfork of the symmetric state  & $\frac{(1-\alpha)(1-3\beta+3\alpha-\alpha\beta)}{4(1-\beta)^2(1+\beta)}$ \\ \hline
        $Hopf$ & Andronov-Hopf of asymmetric mixed state& $\frac{(1-\alpha)(1+2\alpha-2\beta-\beta^2)}{4(1-\beta)^2(1+\beta)}$ \\ \hline
    \end{tabular}
    
\end{table*}
    
\begin{figure}
\centering
\includegraphics[width=0.30\textwidth]{ 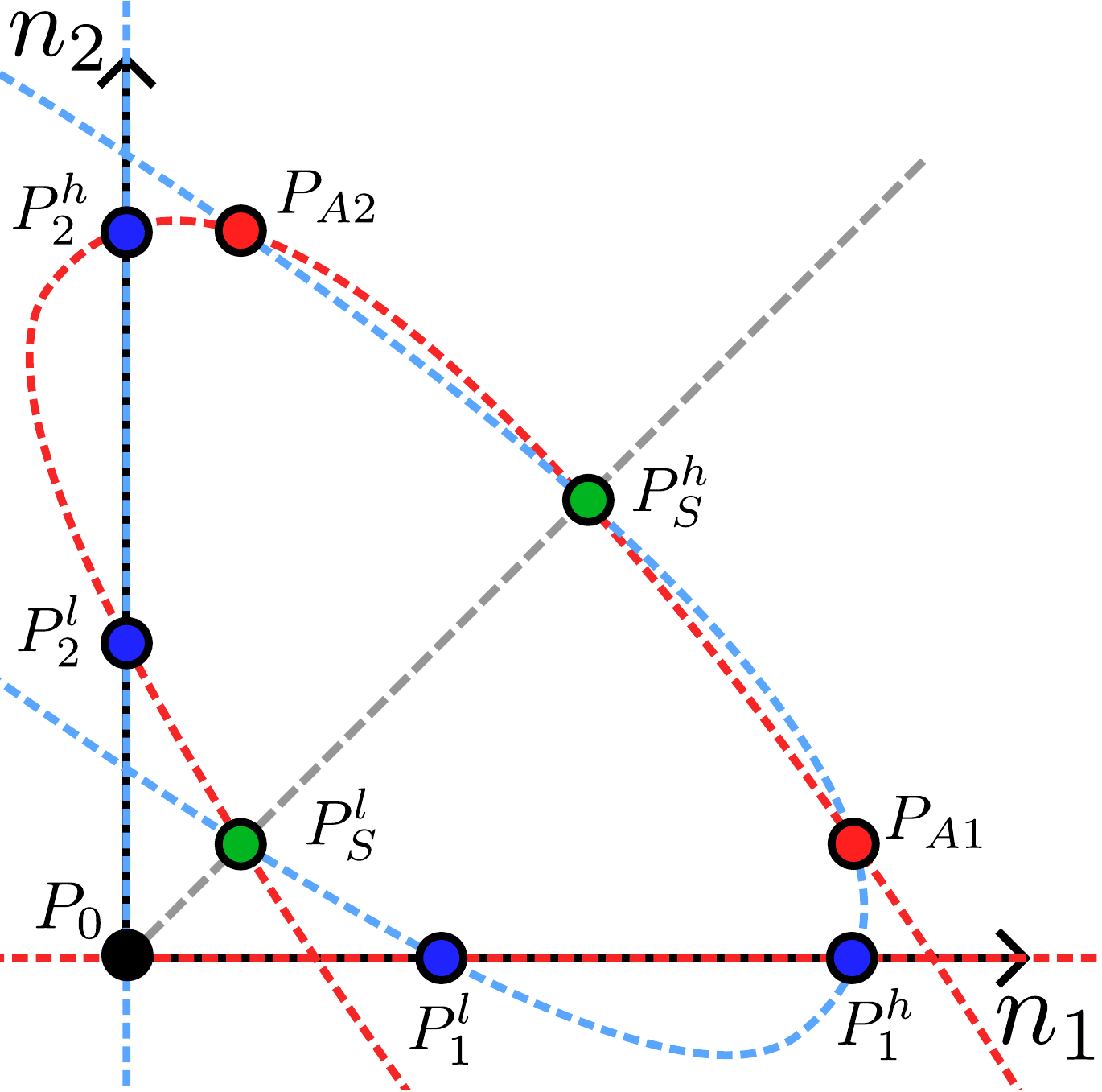}
\caption{Schematic representation of the system's homogeneous steady state (HSS) solutions. The figure shows the nullclines, i.e. zero-growth isoclines for the different species, of the local (homogeneous) system. Nullclines for the $n_1$ ($n_2$) density are shown in blue (red) dashed lines. Points where the two nullclines cross correspond to fixed points of the local systems, i.e. HSSs. There are up to nine of these fixed points, which have been classified into four different groups: unpopulated (black dot), monospecies (blue dots), symmetric mixed states (green dots), and asymmetric mixed states (red dots). Notice the gray dashed symmetry line.}
\label{Fig:Homo_Stdy}
\end{figure}

The local dynamical system can present up to nine different homogeneous steady states (HSS). These fixed points have been classified into four different groups according to the relative concentration of the different species: one unpopulated $P_0$; four mono-species $P_{1}^{l}$, $P_{1}^{h}$, $P_{2}^{l}$, $P_{2}^{h}$; two symmetric mixed $P_S^{l}$, $P_S^{h}$; and two asymmetric mixed $P_{A1}$, $P_{A2}$ (see Fig.~\ref{Fig:Homo_Stdy}). Solutions with a high plant density (labeled with the super-index $h$) and solutions with a lower plant density (labeled with the super-index $l$) can be distinguished in the case of symmetric mixed and monospecies HSSs. These solutions can be related by pairs since they are created via Saddle-Node bifurcations. Due to the symmetry between species, $P_{A1}$, $P_1^l$, and $P_1^h$ have symmetric solutions ($P_{A2}$, $P_2^l$, and $P_2^h$) with interchanged plant densities. In the symmetric case considered here, two symmetric solutions are completely equivalent, so from now on we will drop the sub-indices $1$ and $2$ to refer indistinctly to these solutions, and we will present the results just for the former. 

\begin{figure*}
 \centering
\includegraphics[width=\textwidth]{ 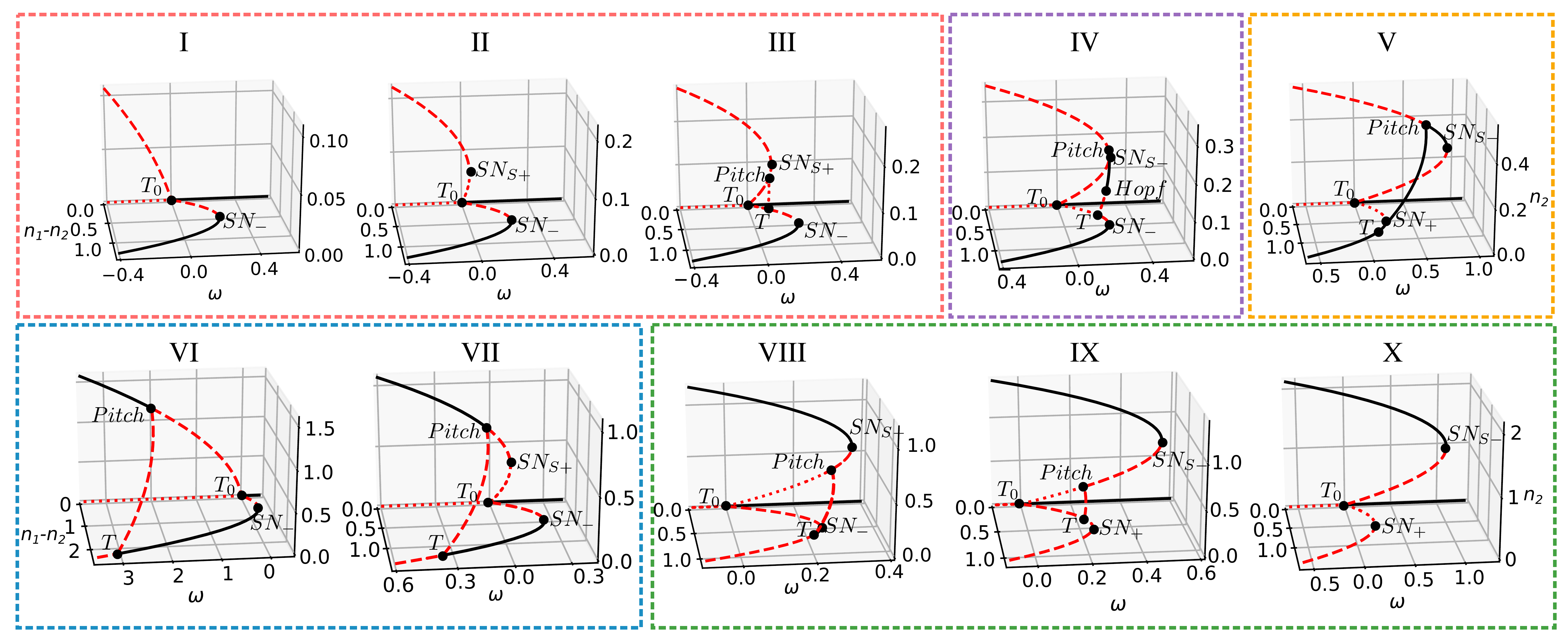}
\caption{\label{Fig:Bif_Diag} Bifurcation diagram as function of the mortality $\omega$ in the different regions shown in Fig.~\ref{Fig:cod_2_phase_diag}. The x-axis represents the difference between the population of the two different species, $n_1-n_2$, the z-axis represents the density of species $2$, $n_2$, and the y-axis represents the control parameter $\omega$.  The branch on the y-axis ($n_2=0$ and $n_1-n_2=0$) corresponds to $P_0$ (black dot in Fig.~\ref{Fig:Homo_Stdy}); branches on the x-y plane ($n_2=0$) correspond to $P^l$ and $P^h$ (blue points in Fig.~\ref{Fig:Homo_Stdy}), branches on the y-z plane ($n_1-n_2=0$) correspond to $P_{S}^{l}$ and $P_{S}^{h}$ (green points in Fig.~\ref{Fig:Homo_Stdy}), and branches out of these planes correspond to $P_A$ (red points in Fig.~\ref{Fig:Homo_Stdy}).  Asymmetric mixed steady states  with a concentration of $n_2$ higher than $n_1$ and monospecies states with species $1$ are not shown but, due to the symmetry of the system, these solutions have the same bifurcation diagram as the equivalent solutions shown here.
Solid black lines represent stable fixed points, red dashed lines saddle points, and red dotted lines unstable nodes or spirals. Colored squares around the diagrams group them into the five different ecological frameworks. The numbers and colors match those used in Fig.~\ref{Fig:cod_2_phase_diag}.}
\end{figure*}

The HSSs are created and change their stability through different bifurcations. Plant density values of each HSS and the corresponding bifurcations are listed in Tables \ref{tab:HSS} and \ref{tab:bif} respectively. In Fig.~\ref{Fig:Bif_Diag} we show the ten qualitatively different bifurcation diagrams of the system as a function of $\omega$. These bifurcation diagrams correspond to values of $\alpha$ and $\beta$ in each corresponding region in Fig.~\ref{Fig:cod_2_phase_diag}.

The unpopulated solution, $P_0$, is a trivial solution of the system which exists for any parameter values. It is stable for $\omega>0$ and unstable for $\omega<0$, losing its stability via a degenerate (due to the imposed symmetry) transcritical bifurcation, $T_0$, at $\omega=\omega_c=0$  involving $P^l$ and either $P_S^l$ or $P_S^h$. The symmetric mixed state involved in this bifurcation is $P_S^h$ for low values of $\alpha$ (Fig.~\ref{Fig:Bif_Diag} I and VI) and $P_S^l$ otherwise (Fig.~\ref{Fig:Bif_Diag} II-V and VII-X). When crossing $T_0$ changing $\omega$, the involved populated solutions change their sign, having biological relevance only those solutions with positive plant density. Note that the positive HSSs involved in the bifurcation are always unstable close to $T_0$ due to dominating low-density intraespecies facilitative interaction.

Monospecies solutions $P^l$ and $P^h$ are characterized by the absence of one of the two species. The system can present four of these solutions, two with the absence of $n_1$ and, equivalently, two symmetric solutions with the absence of $n_2$. These fixed points are generated in two simultaneous monospecific-Saddle-Node ($SN$) bifurcations. The higher branch of the $SN$ corresponds to $P^{h}$, stable under density perturbations of the same species, while the lower branch corresponds to $P^{l}$, which is always unstable.

For a single species the system shows the so-called Allee effect, a positive correlation between the growth rate and the population size for small densities \cite{courchamp2009allee}.
For $\omega>0$, the Allee effect is strong, and there is a threshold (given by $P^l$) below which the plant density decays. The system can show bistability between the unpopulated, $P_0$, and the higher populated monospecific solutions, $P^h$, in this regime. For $\omega<0$, the system displays a weak Allee effect, i.e. there is no threshold for the growth of plant density. Thus, in this regime, $P^h$ will be stable and $P_0$ unstable, while $P^l$ is negative and does not have a biological meaning in this context. The transition between these two regimes, i.e. between monospecific strong and weak Allee effect, occurs through the already discussed transcritical bifurcation $T_0$, involving $P_0$ and $P^l$.

When considering the presence of the other species, the stability of the higher populated monospecific state, $P^h$, is not guaranteed. In regions I-IV and VI-VII, $P^h$ is stable right from the $SN$, which corresponds to a $SN_-$ of the local system. Otherwise, in regions V, IX, and X, the $SN$ corresponds to a $SN_+$, and $P^h$ is unstable to perturbations consisting of a small population of the other species. Away from SN, $P^h$ can still change its stability through a transcritical bifurcation ($T$) with $P_A$, see for instance Fig.~\ref{Fig:Bif_Diag} V and VI-VIII. Crossing this bifurcation point, by decreasing $\omega$, $P_A$ enters a quadrant of negative values, lossing its biological meaning. On the other hand $P^h$ changes its stability, either losing it in a catastrophic transition (see Fig.~\ref{Fig:Bif_Diag} VI-VIII) or gaining it (Fig.~\ref{Fig:Bif_Diag} V).

Symmetric mixed solutions ($P_{S}^{h}$ and $P_{S}^{l}$) are characterized by having the same population of both species, $n_1=n_2$. These solutions are generated at a saddle-node bifurcation with symmetric plant concentrations ($SN_S$).  By decreasing $\omega$ to 0, either $P_{S}^{h}$ or $P_{S}^{l}$, will interact with $P_0$ in $T_0$ changing its sign. In contrast with the monospecific saddle node ($SN$), which occurs always for positive densities, as low densities intraspecific facilitation is assumed in this work, the $SN_S$ might occur for negative population values (see Fig.~\ref{Fig:Bif_Diag} I and VI), and therefore the solutions have no biological meaning at the bifurcation. When this happens (regions I and VI), $P_{S}^{h}$ interacts with $P_0$ at $T_0$, becoming positive for $\omega<0$, while $P_{S}^{l}$ takes always negative values in this case.

When $SN_S$ occurs for positive density values, i.e. in regions II-V and VII-X (see Fig.~\ref{Fig:Bif_Diag}), two different scenarios are found when considering the stability of $P_{S}^{h}$ at the bifurcation point. On one hand, $P_{S}^{h}$ is stable at the bifurcation point in regions IV, V, and VIII-X; where we label the $SN_S$ bifurcation as $SN_{S-}$. On the other hand, $P_{S}^{h}$ is unstable at the bifurcation in regions I-III, VI, and VII; where we denote the $SN_S$ bifurcation as $SN_{S+}$.

A symmetric mixed solution, either $P_{S}^{h}$ or $P_{S}^{l}$, is also involved in a Pitchfork bifurcation ($Pitch$), i.e. a spontaneous symmetry breaking of the system, from where a pair of asymmetric mixed solutions ($P_A$) emerges. Depending on the region, this bifurcation affects one branch or the other of $P_S$ (see Fig.~\ref{Fig:Bif_Diag}). In regions I, II, and X, $Pitch$ involves $P_{S}^{l}$ but for negative values; and $P_A$ does not have biological meaning for any value of $\omega$. In regions III, VIII, and IX, $Pitch$ involves $P_{S}^{l}$ with positive values. In regions IV, V, VI, and VII $Pitch$ affects $P_{S}^{h}$, changing the stability of this point. In this last case, we can make a relevant distinction. In regions IV and V $Pitch$ is supercritical and $P_A$ is stable after the bifurcation, while in regions VI and VII $Pitch$ is subcritical and $P_A$ is unstable. 
Moreover, in region IV, $P_A$ undergoes a Andronov-Hopf bifurcation ($Hopf$), where the stability of $P_A$ changes by decreasing $\omega$ before reaching the $T$ bifurcation. After the Hopf bifurcation a stable homogeneous limit cycle with densities oscillating around $P_A$ is observed. The dynamics of the limit cycle will be discussed in Section \ref{sec:stronglynonlinear}.

The regions in the ($\alpha$, $\beta$) parameter space where each archetypal bifurcation diagram is found are shown in Fig.~\ref{Fig:cod_2_phase_diag}. The curves separating the different regions are given by the projection of codimension-2 bifurcations and singular parameter values of the complete four dimensional parameter space on the ($\alpha$, $\beta$) plane. Regions I and II, and VI and VII are separated by the a codimension-2 bifurcation point in which $SN_S$ and $T_0$ occur for the same parameter values, shown as a red line at $\alpha=-1$ in Fig.~\ref{Fig:cod_2_phase_diag}. Regions II and III, and IX and X are separated by the codimension-2 bifurcation in which $T_0$, $T$, and $Pitch$ occur for the same parameter values, shown as a brown line at $\alpha=1$ in Fig.~\ref{Fig:cod_2_phase_diag}. Regions III and IV, and VII and VIII are separated by the codimension-2 bifurcation in which $Pitch$ and $SN_S$ occur for the same parameter values, marked as a purple line in Fig.~\ref{Fig:cod_2_phase_diag} ($\alpha=\beta$). This codimension-2 point, in the case separating regions III and IV, also involves $Hopf$ and $DH$ bifurcations, in a Bogdanov-Takens bifurcation. The separation between regions IV and V, and VIII and IX are given by the codimension-2 point in which $T$, $Pitch$, and $Hopf$ occur for the same parameter values, marked in green in Fig.~\ref{Fig:cod_2_phase_diag} ($2\alpha=(1+\beta)^2$). Finally, the blue line in Fig.~\ref{Fig:cod_2_phase_diag}, separating regions I, II, and V from VI, VII, and X respectively, represents a singular point in the ($\alpha$, $\beta$) subspace, given by $\beta=1$. Approaching this value of $\beta$, the critical value of $\omega$ at which the bifurcations affecting $P_A$ occur, i.e. $Pitch$ and $T$, diverges to $-\infty$.

\section{Interaction scenarios}

The structure of the HSS bifurcation diagram as a function of the net mortality rate $\omega$ changes depending on the values of inter/intra-species interaction ratios ($\alpha$, $\beta$), as shown in Figs. \ref{Fig:cod_2_phase_diag} and \ref{Fig:Bif_Diag}.
Nevertheless, some of these regimes differ in bifurcations affecting only unstable HSS or involving solutions with negative density values. 
Therefore, we can group the ten cases into just five scenarios with significantly different behavior and ecological interpretation. The regions encompassed in each scenario are shaded with the same color in Fig.~\ref{Fig:cod_2_phase_diag} and grouped by dashed-line boxes in Fig.~\ref{Fig:Bif_Diag}. We further classify the 5 scenarios in two cases: scenarios for large saturation ratios ($\beta > 1$) and scenarios for small saturation ratios ($\beta < 1$).

\subsection{Scenarios for large saturation ratios ($\mathbf{\beta>1}$)}

In this section we study the large-saturation-ratio case, i.e. $\beta >1$, meaning that the interspecific saturation term is larger than the intraspecific one.
Therefore, in this region of the parameter space monospecies meadows are favored, especially for the large densities appearing for small mortality rates. Nevertheless, for intense interspecific facilitation (large values of $\alpha$) stable mixed meadows (either $P_{S}^{h}$ or $P_A$) can appear for  intermediate mortality rates, as well as more exotic behaviors such as oscillations or excitability, due to strongly nonlinear dynamics.

\begin{figure}
\centering
\includegraphics[width=0.5\textwidth]{ 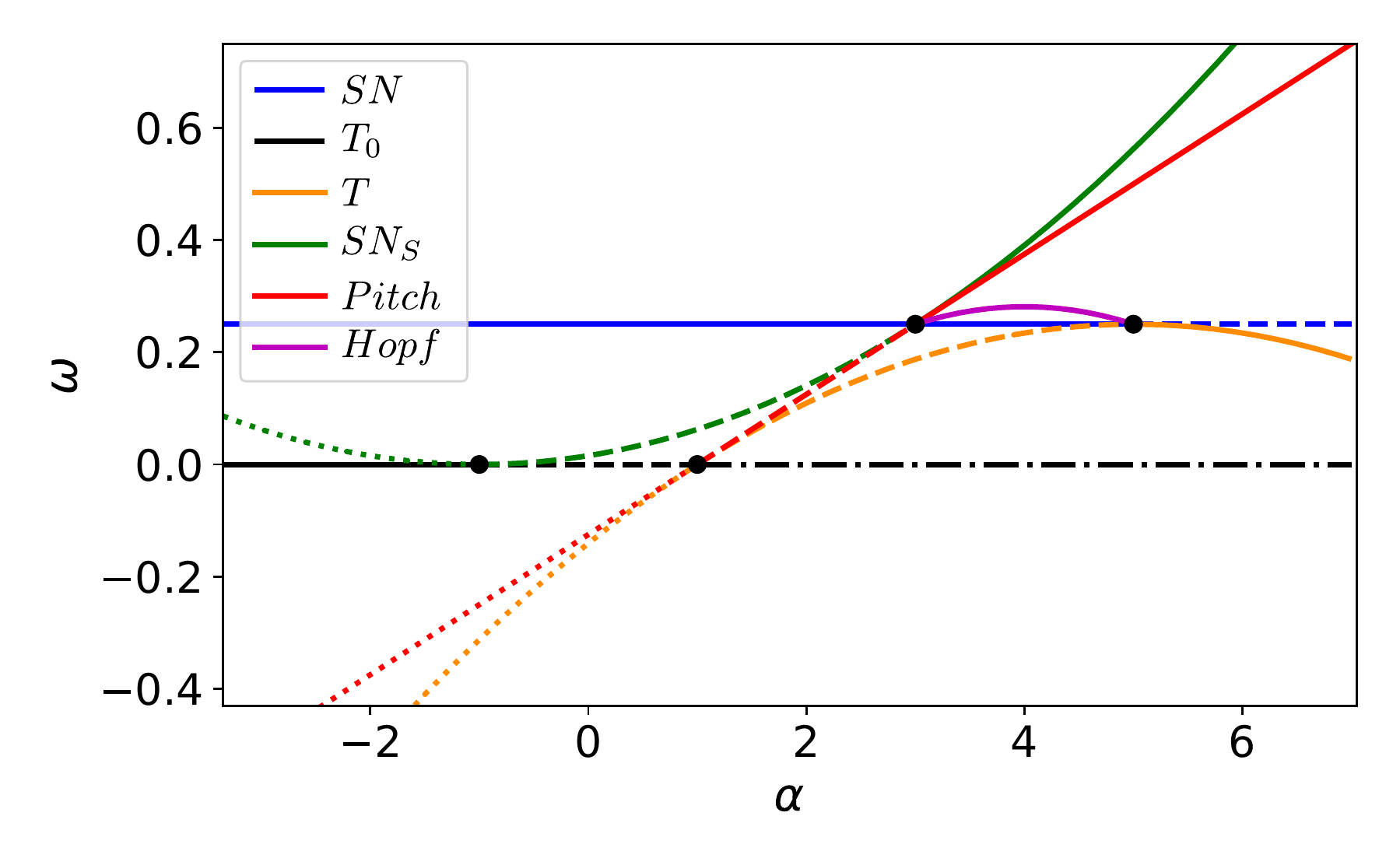}
\caption{($\alpha,\omega$) phase diagram for $\beta = 3$, crossing through regions I-V in Fig. \ref{Fig:cod_2_phase_diag}. This phase diagram is representative of any configuration with $\beta>1$. Dotted lines represent bifurcation involving negative steady points, i.e. solutions without physical meaning. The blue line represents $SN$, solid (dashed) when it corresponds with $SN_-$ ($SN_+$). The green line represents $SN_S$, solid (dashed) when it corresponds with $SN_{S-}$ ($SN_{S+}$). The red solid (dashed) line represents the supercritical (subcritical) pitchfork bifurcation from where $P_A$ emerges. The orange solid (dashed) line represents $T$ involving $P^h$ ($P^l$). The black line represents $T_0$, solid when involving $P_{S}^{h}$, and dashed (dot-dashed) when involving $P_{S}^{l}$ as an unstable node (saddle). Finally, the purple line represents the supercritical $Hopf$ of $P_A$. Dots mark the codimension-2 points.}
\label{Fig:phase_diag_b3}
\end{figure}

A representative phase diagram of this region in the ($\alpha, \omega$) parameter space  is shown in Fig.~\ref{Fig:phase_diag_b3} for $\beta=3$. We next discuss the three different dynamical regimes in this scenario.

\subsubsection{Competitive exclusion scenario}

For $\alpha<\beta$ (regions I, II, and III; pink shaded in Fig. \ref{Fig:cod_2_phase_diag}), there are no stable mixed states. In this scenario, plants compete with each other for all plant densities. Representative bifurcation diagrams are shown in Fig.~\ref{Fig:Bif_Diag} I-III.

For mortality values above the saddle-node bifurcation of the mono-species solutions SN (i.e. $\omega > 0.25$) the only possible state of the system is bare soil ($P_0$), to which any initial condition will converge. For $\omega \in (0,0.25)$ the system shows bistability. On one hand, $P_0$ is still stable, and not dense enough initial conditions die out (strong Allee effect). On the other hand, $P^h$, with either one or the other species, is stable, and dense enough initial conditions will form monospecific meadows. Here $P^l$ acts as a critical density below which the system goes to bare soil and above which the system develops a meadow. For lower moralities ($\omega<0$), the system tends always to monospecific solutions (weak Allee effect).

\begin{figure}
 \centering
\includegraphics[width=0.48\textwidth]{ 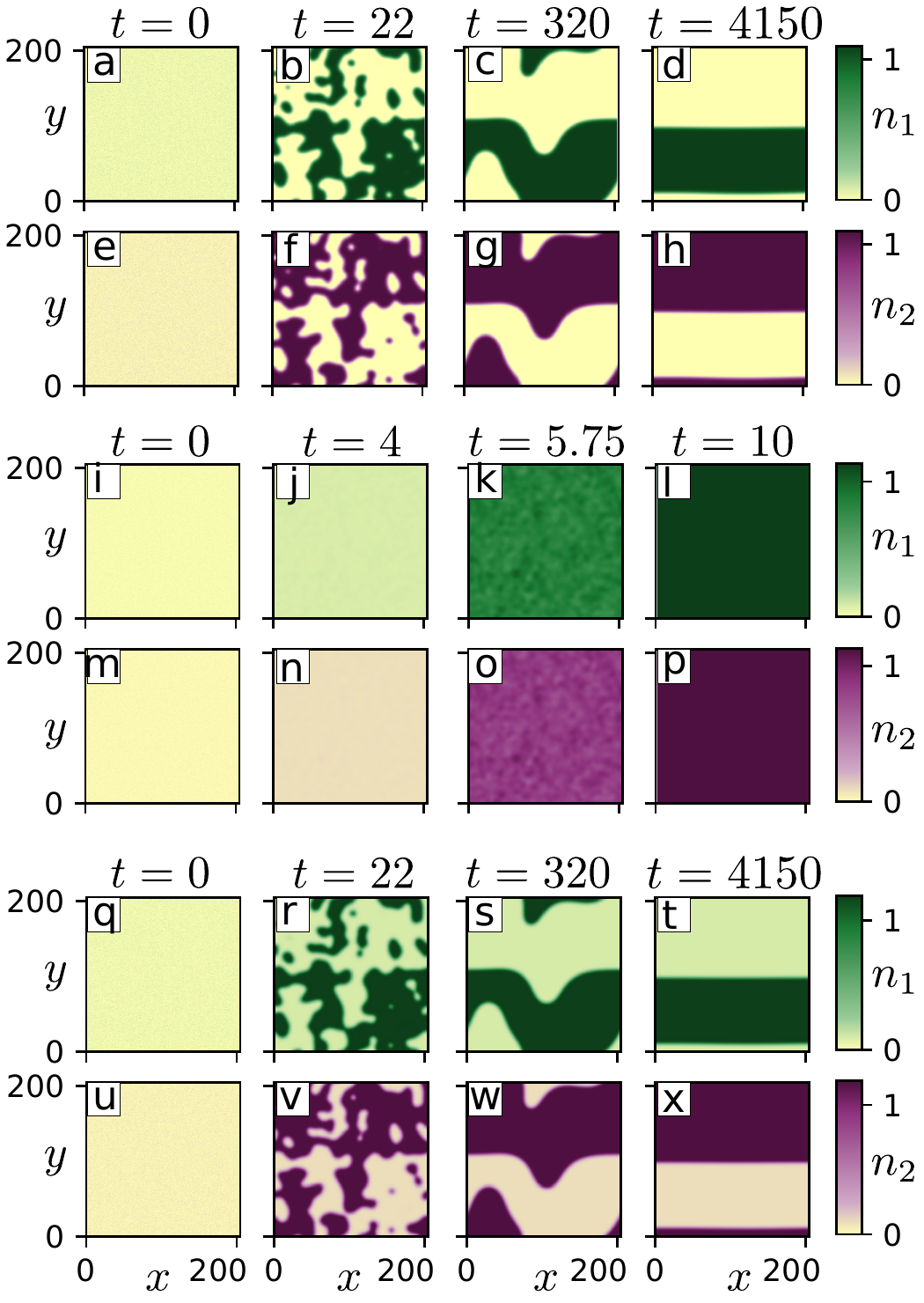}
\caption{\label{Fig:SP-temp-00_} Numerical simulation after the $T_0$, for $\omega=-0.1$. The simulations are initialized around $P_0$ adding small noise. The panels show frames of the density fields for $n_1$ (a-d, i-l, and q-t) and $n_2$ (e-h, m-p, and u-x). The simulation has been performed for three different parameter configurations, showing the 3 different transitions when crossing the bifurcation point. Panels a-h show phase separation to $P^h$, for $\alpha=5$ and $\beta=2.5$. Panels i-p show the transition to $P_S$, for $\alpha=6$ and $\beta=1.5$. Panels q-x show phase separation to $P_A$, for $\alpha=7.7$ and $\beta=2.5$.}
\end{figure}

In this scenario the system displays a hysteresis cycle; the system has a tipping point at $\omega=0.25$ where the populated solution collapses to the bare state. On the other hand, at $\omega=0$, $P_0$ destabilizes and for $\omega < 0$ each species may grow at different places, forming domains separated by fronts. Typically the system shows curvature driven coarsening, in such a way that closed domains will tend to a circular shape and shrink, following its size a $t^{1/2}$ scaling law, until disappearing completely \cite{Gomila01}. In this case, the final state at long times is always either a single species meadow or regions of different species separated by flat fronts Fig.~\ref{Fig:SP-temp-00_} a-h.
This phase separation scenario can be related to dominating competitive ecological interactions between species. This situation is structurally unstable, and any breaking of the symmetry between species will make the dominant one to overrun the other and colonize all the space.

\subsubsection{Strongly nonlinear regime}
\label{sec:stronglynonlinear}

The region with $\beta<\alpha<\frac{1}{2}(1+\beta^2)$ (region IV, shaded in purple in Fig. \ref{Fig:cod_2_phase_diag}) presents a highly nonlinear behavior for intermediate values of mortality. This behavior is generated due to the interplay between strong quadratic interspecies facilitation terms and also strong cubic interspecies saturation.

\begin{figure}
 \centering
\includegraphics[width=0.48\textwidth]{ 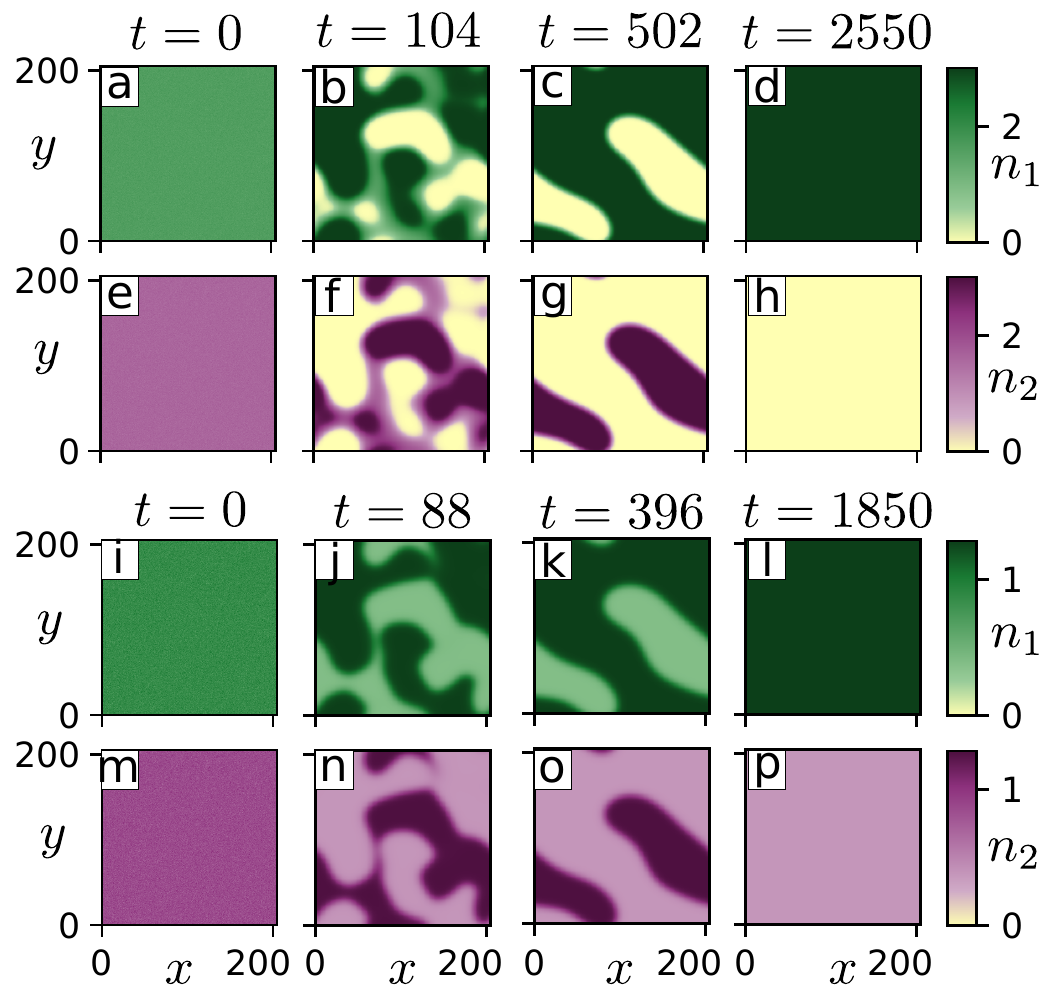}
\caption{\label{Fig:SP-temp-pitch_} Numerical simulations close to $Pitch$ bifurcation. The simulations are initialized around $P_{S}^{h}$ with small gaussian noise. The panels show frames of the density fields for $n_1$ (a-d and i-l) and $n_2$ (e-h and m-p). The simulation has been performed for two different parameter configurations, one showing the behavior after the subcritical $Pitch$ (a-h), with $\omega=-6.04$, $\alpha=-2$, and $\beta = 0.3$; and the other after the supercritical $Pitch$ (i-p), with $\omega=-0.616$, $\alpha=6$, and $\beta = 2$.}
\end{figure}

As usual, for large enough mortality rates, the only possible final state of the system is $P_0$, and any initial non-zero population decays. For lower mortality values, the system shows bistability between $P_0$ and $P_{S}^{h}$. However, for smaller mortality values, $P_{S}^{h}$ destabilizes through a supercritical pitchfork bifurcation, leading to a phase separation dynamics of the two asymmetric solutions $P_A$, as shown in Fig.~\ref{Fig:SP-temp-pitch_} i-p. For even lower mortalities, $P_A$ undergoes a Hopf bifurcation and densities $n_1$ and $n_2$ oscillate around these states. 

\begin{figure}
    \centering
    \includegraphics[width=0.48\textwidth]{ 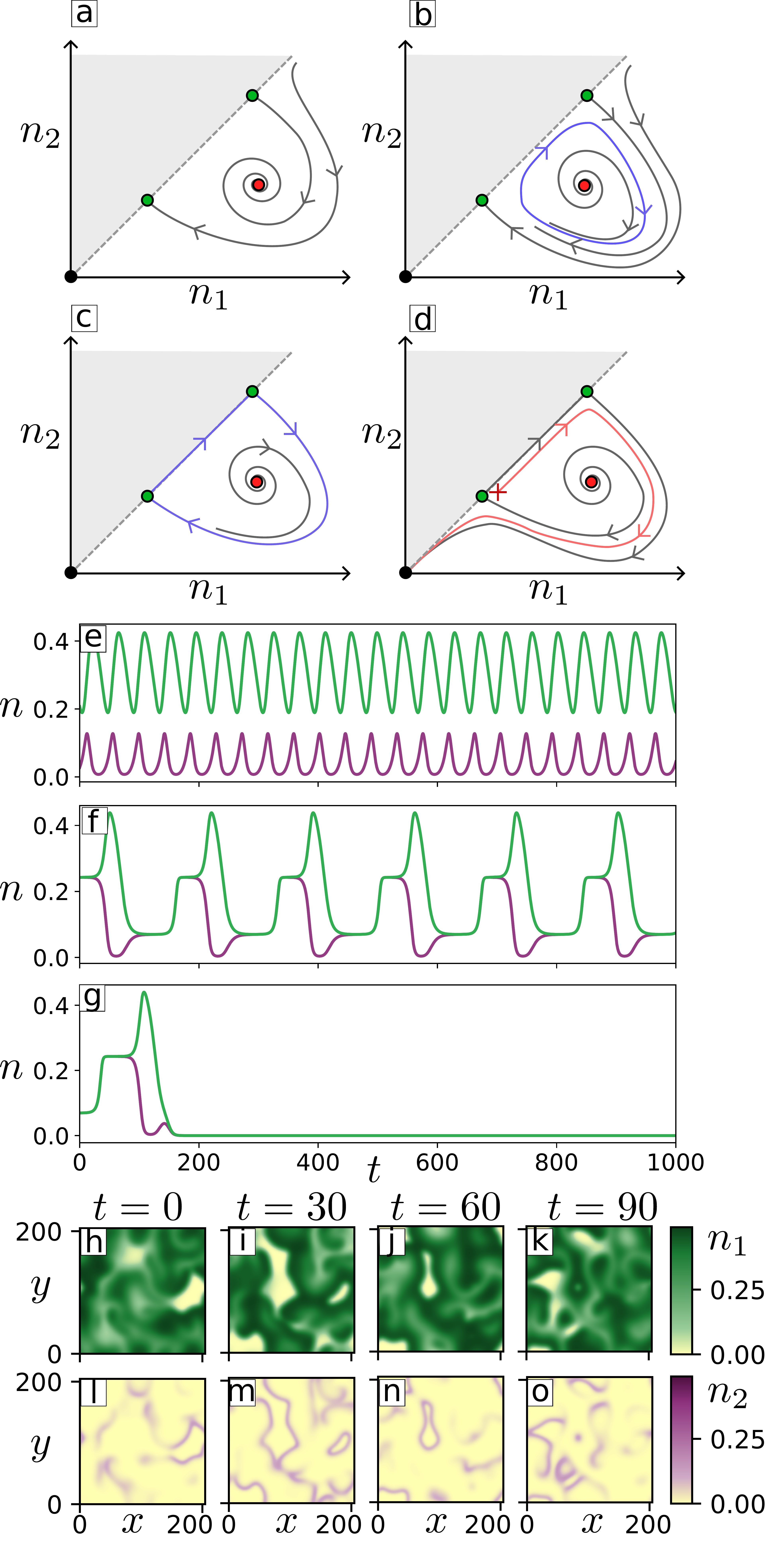}
    \caption{ Oscillatory and turbulent regimes around asymmetric solutions. Panels a)-d) show sketches of the phase diagram for different values of $\omega$ in region V: a) between $Pitch$ and $Hopf$; b) after the $Hopf$, $P_A$ destabilizes and a stable limit cycle emerges. Decreasing $\omega$ further, the limit cycle grows in amplitude and approaches $P_{S}^{h}$ and $P_{S}^{l}$, until it touches them, c), at a double heteroclinic ($DH$). Approaching this bifurcation point, the period of the oscillations diverges and the limit cycle is destroyed after crossing it.     
    After this point, the system shows local excitability, see red trajectory in panel d). Panels e)-g) show the time evolution of each species density in the oscillatory regime for $\alpha=4$ and $\beta=3$, and (e) $\omega=0.275$ far from the $DH$ bifurcation, (f) $\omega=0.27082$ closer to $DH$, and (g)  $\omega = 0.27$ in the excitable regime passed the $DH$. Panels h)-o) show the turbulent regime observed in the excitable region for $\omega = 0.251$.}
    \label{fig:Het_Bif}
\end{figure}

The dynamics of the limit cycle for decreasing values of $\omega$ is shown in Fig.~\ref{fig:Het_Bif}. Decreasing $\omega$, the limit cycle growths in amplitude and approaches $P_{S}^{l}$ and $P_{S}^{h}$ simultaneously (see Fig.~\ref{fig:Het_Bif}a, b and c). Close to these fixed points, the limit cycle slows down (see Fig.~\ref{fig:Het_Bif}e and f). Eventually, decreasing $\omega$ even more, the limit cycle touches $P_{S}^{l}$ and $P_{S}^{h}$ in a Double-Heteroclinc connection ($DH$), as shown in Fig.~\ref{fig:Het_Bif} c). After this bifurcation point, the limit cycle is destroyed and the local system presents Type-I excitable behavior (see Fig.~\ref{fig:Het_Bif}d and g).

In this excitable regime, homogeneous initial conditions below a threshold, given by the stable manifold of $P_{S}^{l}$, decay to the bare state. Homogeneous initial conditions above this threshold will, however, make a large excursion in phase space to finally come back to the bare state  again, an excitable trajectory (see Fig.~\ref{fig:Het_Bif}d and g). This leads to the apparent paradoxical absence of persistent populated solutions in the so called "excitable regime". This paradoxical behavior can be related to the "enrichment paradox" \cite{Rosenzweig1971} observed in many population dynamics models. However, in this case, localized initial conditions grow in this regime, creating a turbulent state that expands onto bared soil. An example of this regime is shown in Fig.~\ref{fig:Het_Bif} h-o. At difference with other models \cite{Arinyo2021,Moreno2022}, for the parameters used in this study we have not observed stable travelling pulses in the excitable regime, only turbulent states.

\subsubsection{Obligate mutualism to monospecific transition}

For $\alpha>\frac{1}{2}(1+\beta^2)$ (region V, shaded in yellow in Fig. \ref{Fig:cod_2_phase_diag}) we observe a smooth transition from mixed symmetric, $P_{S}^{h}$, to monospecies, $P^h$, meadows through asymmetric states, $P_A$ (see panel V in Fig. \ref{Fig:Bif_Diag}). This transition can be understood as an obligate mutualism interaction for low plant densities, but a competitive interaction for high densities, giving a competitive exclusion scenario for small mortalities.

In this scenario, the system presents a hysteresis cycle between populated solutions and $P_0$. For high mortality, $P_0$ is the only possible state. Decreasing the mortality the system crosses $SN_{S}$, after which it shows bistability between $P_0$ and $P_{S}^{h}$. If we follow the populated branch $P_{S}^{h}$ while decreasing $\omega$, there is a point where  the system eventually crosses a supercritical pitchfork bifurcation ($Pitch$) and $P_{S}^{h}$ losses stability. Initial conditions around $P_{S}^{h}$ slightly below this point tend to phase separate driven by curvature, forming domains of either one of the two assymmetric solutions, $P_{A1}$ or $P_{A2}$, as shown in Fig.~\ref{Fig:SP-temp-pitch_} i-p. Decreasing $\omega$ even more, $P_A$ becomes more and more asymmetrical, one of the two species increasing its density while the other decreases it until $P_A$ eventually reaches $P^h$ in $T$. This gives a continuous transition of the populated stable solutions from $P_{S}^{h}$ to $P^h$ while decreasing $\omega$.

Here the stable bare state coexists with stable populated solutions (either $P_{S}^{h}$, $P_A$ or $P^h$) until $\omega=0$ where it losses its stability through $T_0$. Crossing this threshold, the system undergoes a phase separation involving either the monospecific solutions (see Fig.~\ref{Fig:SP-temp-00_} a-h) or the asymmetric mixed solutions (see Fig.~\ref{Fig:SP-temp-00_} q-x),  depending on the relative position of $Pitch$ and $T$ due to the parameter values.

\subsection{Scenarios for small saturation ratio ($\beta<1$)}

In this section we study the scenarios with a small saturation ratio, i.e. when $\beta<1$ and therefore the intraspecies saturation is greater than the interspecies one. In this region of parameter space, the $P_{S}^{h}$ is favored, especially for small mortality values where it shows large densities. Nevertheless, for small values of the parameter $\alpha$, describing interspecies competition or just very weak interspecies facilitation for low plant densities, the system can also show monospecific meadows for intermediate mortalities.

\begin{figure}
\centering
\includegraphics[width=0.5\textwidth]{ 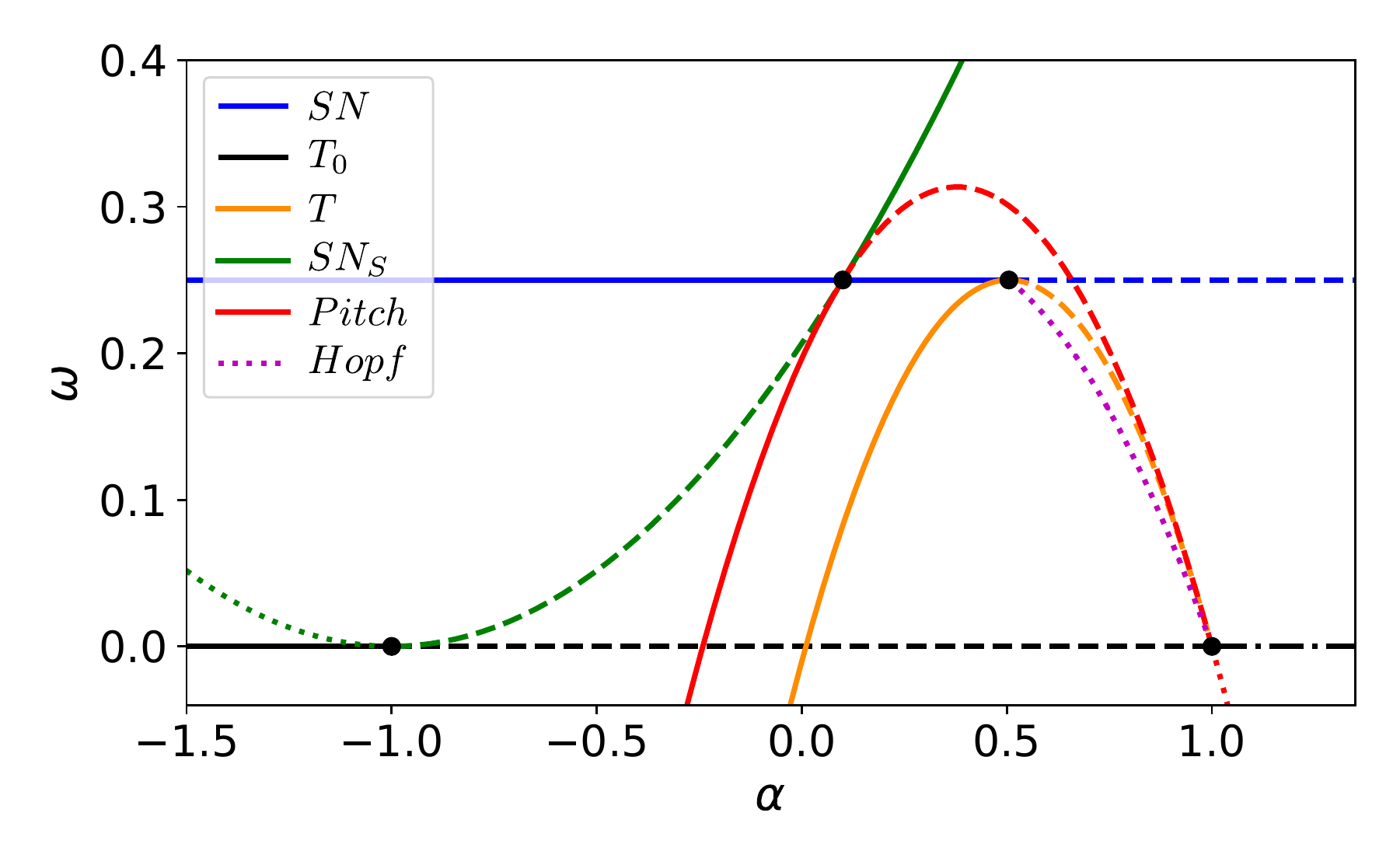}
\caption{Phase diagram for $\beta = 0.1$ crossing through regions VI-X. This phase diagram is representative of any configuration with $\beta<1$.
Dotted lines represent bifurcation involving negative steady points, i.e. solutions without physical meaning.
The blue solid (dashed) line represents the $SN_-$ ($SN_+$) bifurcation. 
The green line represents the $SN_S$, solid (dashed) when corresponding to  $SN_{S-}$ ($SN_{S+}$).
The red solid (dashed) line represents the subcritical (supercritical) $Pitch$ involving $P_{S}^{h}$ ($P_{S}^{l}$).
The orange solid (dashed) line represents $T$ involving $P^h$ ($P^l$).
The black line represents $T_0$, solid when involves $P_{S}^{h}$, dashed when involves $P_{S}^{l}$ while it is a node, and dotted-dashed when involves $P_{S}^{l}$ while it is a saddle.
The purple dotted line represents the $Hopf$ bifurcation of $P_A$, with negative density values.
Dots mark the codimension-2 points.}
\label{Fig:phase_diag_b01}
\end{figure}

A representative phase diagram of this region in the ($\alpha, \omega$) parameter space  is shown in Fig.~\ref{Fig:phase_diag_b01} for $\beta=0.1$. We next discuss the two different cases in this scenario.

\subsubsection{Competitive exclusion to facultative mutualism transition}

For small values of $\alpha$, i.e. $\alpha < \beta$ (regions VI and VII, blue shaded in  Fig. \ref{Fig:cod_2_phase_diag}), the system tends to $P^h$ for intermediate mortality values, and to $P_{S}^{h}$ for lower mortalities. The transition between these two configurations is abrupt, and the system shows a hysteresis cycle. This hysteresis cycle coexists with another one between populated and unpopulated solutions (see Fig.~\ref{Fig:Bif_Diag} VI-VII).

The transition from monospecific, $P^h$, to mixed symmetric, $P_{S}^{h}$, meadows occurs through $T$, that involves the (unstable) $P_A$. The transition from $P_{S}^{h}$ to $P^h$ occurs after a subcritical pitchfork, $Pitch$. This transition leads to a spontaneous symmetry breaking and a phase separation of the two monoespecific solutions (see Fig.~\ref{Fig:SP-temp-pitch_} a-h).

The populated-unpopulated hysteresis cycle involves $P^h$, which is destroyed at $SN$ for $\omega = 0.25$. For larger mortalities, any initial condition decays to the bared state $P_0$. $P_0$ is stable above $T_0$. After this bifurcation, and depending on the parameters, the system can show a phase separation to $P^h$ (see Fig.~\ref{Fig:SP-temp-00_} a-h) or converge to $P_{S}^{h}$ (see Fig.~\ref{Fig:SP-temp-00_} i-p). 

\subsubsection{Obligate and facultative mutualism}

For large values of $\alpha$, i.e. $\alpha >\beta$ (regions VIII, IX, and X, shadowed in green in  Fig. \ref{Fig:cod_2_phase_diag}), $P_{S}^{h}$ is always stable below $SN_s$. Representative bifurcation diagrams of this region are shown in Fig.~\ref{Fig:Bif_Diag} VIII, IX and X.

For large mortality values, the symmetric mixed solution has a tipping point ($SN_s$) and the system collapses to the bare state. This bare state coexists with the symmetric mixed solution until $\omega=0$, after which the system converges to the symmetric mixed solution (see Fig.~\ref{Fig:SP-temp-00_} i-p). 

In region VIII there is a small set of mortality values for which the monospecific solution is also stable, showing the system bistability between monospecific and symmetric mixed populated states.

\section{Conclusions}
We have presented a general spatiotemporal population dynamics model for two interacting seagrass species. 
The interaction between species has been introduced as a coupling through the mortality rate, with up to quadratic density dependent terms. This allows modeling different types of interactions. Regarding intraspecific interactions, these nonlinear terms allow low-density facilitation and high-density sturation leading to bounded solutions, i.e. Allee effect. For the interspecific interactions, the nonlinear density dependence allows, for some parameters, the prevalence of monospecific solutions, and species segregation for large density solutions associated with low mortality rates. The system include also non-linear diffusion and a gradient squared term to model clonal reproduction.

In this work we have analyzed in detail the symmetric scenario of the general model, where intraspecific interactions are equal for both species and interspecific interaction is reciprocal. This scenario reduces the model parameters to just 4 in its adimensional form. We have characterized the bifurcation diagram of the symmetric scenario, which can be considered as a backbone of the complete system.

The parameter space of the symmetric scenario can be divided into ten different regions according to the values of the biotic parameters $\alpha$ and $\beta$ determining the ratio between the intraspecific and interspecific interaction strengths. The bifurcation diagrams of the fixed points in each of these regions as a function of the net mortality rate $\omega$, the parameter depending on abiotic factors, are qualitatively different. Furthermore, we can group these regions into five different scenarios with different ecological interpretations, including obligate and facultative mutualism, competitive exclusion, and strongly nonlinear regimes, as well as transitions between them.  

Some of these scenarios (regions VI-X in Fig. \ref{Fig:cod_2_phase_diag}) are compatible with a linear interaction between species, corresponding in the model to $\beta =0$. Nevertheless, some of the dynamics found in regions I-V are incompatible with just lineal interspecific interaction in a symmetric system. These dynamics include stable asymmetric states, oscillations, turbulence, and competitive exclusion.

We have only studied in detail the symmetric case of the proposed models. Nevertheless, in many real cases, the interacting species are very different and the interaction can be asymmetric. Therefore, a natural extension of our work is to apply the model to particular cases, as has been already done with some seagrass microscopic models and macroscopic single species systems \cite{Llabr2022a,Llabr2022b,Ruiz-Reynes2017}.

\appendix
\section{Linear stability analisis.}

In this appendix, we describe the stability analysis used to study the bifurcations affecting HSSs. In particular, we show that there are no finite wavelength instabilities, a.k.a. Turing instabilities, for any of these solutions.

To study the linear stability of HSSs we consider small perturbations of the form:
\begin{align}
    \vec{n}_q=\vec{n}_q^{0} e^{\sigma_q t +i q x}
\end{align}
where  $\sigma _q$ is the eigenvalue associated with the eigenvector $\vec{n}_q^0$ of the Jacobian matrix around the HSS:
\begin{equation}
    J_q(n_1^*,n_2^*)= J_0(n_1^*,n_2^*)+
    \begin{pmatrix}
    -(1+\delta n_1^*)q^2 & 0\\
    0 & -(1+\delta n_2^*)q^2 
    \end{pmatrix}
    \label{Jacobian_q}
\end{equation}
where $J_0$ is the homogeneous Jacobian matrix, given by:
\begin{align}
    J_0(n_1^*,n_2^*)&=
    \begin{pmatrix}
    J_{11} & J_{12}\\
    J_{21} & J_{22} 
    \end{pmatrix} =
    \begin{pmatrix}
    Q(n_1^*,n_2^*)+n_1^*-2n_1^*(n_1^*+\beta n_2^*) & n_1^*[\alpha-2\beta(n_1^*+\beta n_2^*)]\\
    n_2^*[\alpha-2\beta(n_2^*+\beta n_1^*)] & Q(n_2^*,n_1^*)+n_2^*-2n_2^*(n_2^*+\beta n_1^*) 
    \end{pmatrix}
\end{align}

The bifurcations presented in this paper can straightforwardly be obtained through the study of the eigenvalues of the $J_0$ matrix.

To detect pattern forming instabilities one must consider the full Jacobian $J_q(n_1^*,n_2^*)$. Although in the symmetric case the diffusion coefficients are equal, $d_{10}=d_{20}$, the presence of nonlinear diffusion does not allow to discard, a priory, the presence of a Turing instability in the system. In what follows, however, we prove that, despite nonlinear diffusion, no Turing instability can take place in the symmetric case for any of the HSSs.

Six different conditions must be fulfilled in order to a Turing instability to take place. First, both field of the homogeneous solution must be positive to have physical meaning:
\begin{equation}
    n_{1,2}^* \geq 0.  \label{A.Cond.1}
\end{equation}
Second, the solution might be linearly stable under homogeneous perturbation, and therefore following two conditions must be fulfilled:
\begin{align}
    \tau &= J_{11} + J_{22} < 0 \label{A.Cond.2}\\
    \Delta &= J_{11}J_{22}-J_{12}J_{21}>0. \label{A.Cond.3}
\end{align}
Finally, the transition must happen for a real critical wavenumber and a positive value of the control parameter $\delta$:
\begin{align}
    q^2_c&=\frac{J_{11}(1+\delta n_2^*)+J_{22}(1+\delta n_1^*)}{2 (1+\delta n_1^*)(1+\delta n_2^*) } >0 \label{A.Cond.4}\\
    \delta_c&>0. \label{A.Cond.5}
\end{align}

\subsection{ Turing of the unpopulated solution}
In this subsection, we prove that the unpopulated solution has no physically meaningful Turing instability.

First, the growth of a non-zero-wavenumber perturbation on top of the bare state implies regions of the space with a negative value of the population density of at least one species. These solutions, therefore don't have physical meaning and are forbidden, by construction, on the system.

Nevertheless, we can compute the square critical wavenumber $q_c^2$ equation to obtain, $q_c^2 = -\omega$. Therefore this critical wavenumber only exists for negative values of $\omega$. Computing the determinant and the trace of the linearized system for perturbations with $q_c$, we obtain $\tau_c=0$ and $\Delta_c=0$, showing that the value at this point does not depend on $\delta$, meaning that the eigenvalue associated with $q_c$ will be a double geometric-degenerated zero and never will be positive. This point, therefore, is not associated with a Turing instability but is a consequence of the symmetries of the problem.

\subsection{Turing of the monospecies solutions}
In this subsection, we prove that there is no physically meaningful Turing instability for monospecific homogeneous solutions. As the matrix given by Eq.~(\ref{Jacobian_q}) is triangular, the eigenvalues are easily obtained for the $P^{l,h}$. Their eigenvalues are given by the following equation:\begin{align}
    \lambda_1(q) = n^* -2n^{*2} -(1+\delta n^*) q^2 \nonumber \\
    \lambda_2(q) = -\omega + \alpha n^* -\beta ^2 n^{*2} -q^2
\end{align}
where $n^*$ is the plant density of the populated specie. The only relative maximum of both eigenvalues is for $q=0$ and, therefore, no Turing instability can take place for monospecific HSSs. 

\subsection{Turing of the symmetric mixed solution}
Using Eq.~(\ref{A.Cond.2}) and Eq.~(\ref{A.Cond.4}) for $P_S^{l,h}$ we obtain:
\begin{align}
    q_c^2 = \frac{\tau}{4(1+\delta n_i^*)}.
\end{align}
Assuming conditions (\ref{A.Cond.1}), and (\ref{A.Cond.5}) are fulfilled, we obtain that $q_c^2>0 \iff \tau>0$, which contradicts either condition (\ref{A.Cond.2}) or (\ref{A.Cond.4}). Therefore it is not possible to fulfill all the conditions at the same time and there is no Turing instability for symmetric mixed solutions. 

\subsection{ Turing of asymmetric mixed solutions}

To work with asymmetric mixed solutions we will make use of the following change of variables: $\mu=n_1+n_2$, $\nu=n_1-n_2$. The asymmetric mixed solution is given by $\mu^*=\frac{1-\alpha}{1-\beta^2}$ and $\nu^*=\pm2\sqrt{\frac{\omega_P -\omega}{(1-\beta)^2}}$. Notice that condition (\ref{A.Cond.1}) is only fulfil when $|\nu^*|<\mu^*$ and $\mu^*>0$.

With this change of variables, condition (\ref{A.Cond.2}) reads:
\begin{align}
    \tau=\mu^*(1-\mu^*-\beta \mu^*)-\nu^{*2}(1-\beta) <0.
    \label{Cond_Asym_Turing}
\end{align}
While, assuming conditions (\ref{A.Cond.1}) and (\ref{A.Cond.5}) are fulfilled, we can focus just in the numerator of (\ref{A.Cond.4}) and rewrite it as:
\begin{align}
    \tau + \delta(J_{11}n_2^*+J_{22}n_1^*)>0.
\end{align}
As $\tau<0$, a necessary but not sufficient condition for this last inequality is:
\begin{align}
    J_{11}n_2^* + J_{22}n_1^* = \frac{1}{2}(\mu^{*2}-\nu^{*2})(1-\mu^*-\beta \mu^*)>0,
\end{align}

and,as $|\nu^*|<\mu^*$, this condition reduces to $1-\mu^*-\beta \mu^* >0$, or, as $\beta>0$, to $\mu^*<\frac{1}{1+\beta}$.

Substituting this last expression on (\ref{Cond_Asym_Turing}), assuming $\mu^*>0$, we arrive to the necessary condition $\beta<1$. Now, from the same expression and considering again $|\nu^*|<\mu^*$ we arrive to to:
\begin{align}
    \mu^*(1-\mu^*-\beta \mu^*)<\nu^{*2}(1-\beta)<\mu^{*2}(1-\beta)
\end{align}
and therefore:
\begin{align}
    \mu^*>\frac{1}{2}.
\end{align}

Altogether we get $\frac{1}{2}<\mu^*<\frac{1}{1+\beta}<\frac{1}{2}$, wich has no solution. Therefore we conclude that there is no Turing instability of the asymmetric solutions.

\begin{acknowledgement}
We acknowledge financial support from project CYCLE (PID2021-123723OB-C22) funded by MCIN/AEI/10.13039/501100011033 and ERDF “A way of making Europe”, the María de Maeztu project CEX2021-001164-M funded by the MCIN/AEI/10.13039/501100011033, and  the European Union’s Horizon's 2020 research and innovation programme (Grant agreement ID: 101093910, Ocean Citizen). PMS acknowledges support from the FPI grant RTI2018-095441-B-C22.
\end{acknowledgement}


\end{document}